\documentclass[a4paper,12pt]{article}
\usepackage{jheppub}
\usepackage[T1]{fontenc}
\usepackage[utf8]{inputenc}
\usepackage{float}
\usepackage{slashed}
\usepackage{multicol,multirow}
\usepackage{amssymb}
\usepackage{amsmath}
\usepackage{subcaption}
\usepackage{array}
\usepackage{color}
\usepackage{hyperref}
\hypersetup{colorlinks=true}
\usepackage{longtable}
\usepackage{graphics,graphicx,epsfig,ulem,booktabs}
\numberwithin{equation}{section}

\definecolor{blue}{rgb}{0.2, 0.4, 1.0}
\definecolor{green}{rgb}{0.1,0.8,0.2}
\definecolor{orange}{rgb}{0.95,0.45,0.0}
\definecolor{cyan}{rgb}{0.0,0.75,0.8}

\newcolumntype{C}[1]{>{\centering\let\newline\\\arraybackslash\hspace{0pt}}m{#1}}

\newcommand{\tud}{\tau (B^+)/\tau (B_d)}

\newcommand{\Oq}{{\cal O}}

\newcommand{\B}{{\cal B}}

\title{
\begin{center}
\boldmath 
Constraining $B$-Mesogenesis models with inclusive and exclusive decays.
\end{center}
}

\preprint{P3H-24-099,SI-HEP-2024-30}

\author[a]{Alexander Lenz,}
\author[a]{Ali Mohamed,}
\author[a]{Zachary Wüthrich}

\affiliation[a]{Physik Department, Universit\"{a}t Siegen, 
Walter-Flex-Str. 3, 57068 Siegen, Germany}

\emailAdd{alexander.lenz@uni-siegen.de}
\emailAdd{Ali.Mohamed@uni-siegen.de}
\emailAdd{Zachary.Wuethrich@uni-siegen.de}

\abstract{The  $B$-Mesogenesis model explains the
matter-antimatter asymmetry 
and leads to the right amount of dark matter in the Universe. 
In particular, this model predicts new decay channels of the $b$ quark.
We investigate the modification of inclusive $b$-hadron decay rates and of the lifetimes of different $B$
mesons due to these new decay channels and compare our results with available predictions for exclusive $B$ meson decays.
We find a small surviving parameter space 
where the  $B$-Mesogenesis model is working and which has not been excluded by experiment.
Experimental investigations in the near future should be able to test this remaining parameter space and thus either exclude or confirm the  $B$-Mesogenesis model.
}

\begin{document}

\maketitle

\section{Introduction}
The  Standard Model (SM) of particle physics is extremely
successful in describing the microscopic world, it
leaves, however, several fundamental questions
unanswered, like the existence of  matter in the Universe and the presence of dark matter.
The $B$-Mesogenesis model discussed in
Ref.~\cite{Alonso-Alvarez:2021qfd} presents an 
interesting framework to solve both of these issues. In particular, this model predicts new decay channels of the $b$-quark into SM baryons and dark (i.e. invisible in particle physics detectors) anti-baryons.   
These new decay channels lead to new exclusive decay channels of $b$ hadrons, as well as to a modification of inclusive decay channels and therefore of the lifetimes and to a modification of mixing observables.  
\\
For the new exclusive decay channels of $b$ hadrons, theory predictions have been made recently within the framework of light cone sum rules \cite{Balitsky:1989ry,Khodjamirian:2023wol}:
$B^+$-meson decays into a proton $p^+$ and a dark antibaryon $\bar{\psi}$, i.e.  $B^+ \to p^+ \bar{\psi}$,  have been studied in 
Ref.~\cite{Khodjamirian:2022vta} and higher twist corrections to this decay have been estimated in
Ref.~\cite{Boushmelev:2023huu}.
In these two works the $B^+$ meson was described by an interpolating current, while the nucleon was described by its distribution amplitudes, which are known to higher twist \cite{Braun:2001tj,Lenz:2003tq,Braun:2006hz,Lenz:2009ar}.
$B$-meson decays into different baryons ${\cal B}$
(octet baryons or charmed anti-triplet baryon) and dark antibaryon,  $B^+ \to {\cal B} \bar{\psi}$, 
were investigated in Ref.~\cite{Elor:2022jxy} using the leading twist distribution amplitudes of the different light baryons from 
Ref.~\cite{RQCD:2019hps} and some simplifying assumptions for the charmed baryons.
Finally, decays of heavy $\Lambda_b$ and $\Xi_b$
baryons into a pseudoscalar meson and a dark baryon were considered in Ref.~\cite{Shi:2024uqs}, using the $\Lambda_b$ distribution amplitudes obtained in
Refs.~\cite{Ball:2008fw,Duan:2022uzm}.
\\
In this paper we will determine the full inclusive decay rate $\Gamma (b \to d u \psi)$\footnote{Semi-inclusive decay rates were calculated in Ref.~\cite{Shi:2023riy}.}. To fullfill the requirement of the $B$-mesogenesis model, the corresponding inclusive branching fraction has to have a value of at least $10^{-4}$. By requiring this bound to be fullfilled, we will get lower bounds on the elementary $b \to X \psi$ couplings, which can be insert into the predictions of the exclusive decays in order to get lower bounds on the exclusive branching fraction, which are close to the  experimental prospects for detecting these channels.
\\
Lifetimes of $b$ hadrons and lifetime ratios like  $\tud$ are by now experimentally known with a high precision~\cite{Amhis:2022mac}
\begin{equation}
   \left(\frac{\tau (B^+)}{\tau (B_d)}\right)^{\rm Exp.}
    = 1.076 \pm 0.004 \, .
\label{eq:ratio_Exp}
\end{equation}
On the theoretical side, predictions for this ratio can be obtained within the framework of the heavy quark expansion (HQE), which has proven to be a powerful method to perform systematic studies of inclusive decay widths of heavy hadrons, see e.g.\ the reviews~\cite{Lenz:2014jha,Albrecht:2024oyn}.
Based on the calculations in Refs.~\cite{King:2021jsq,King:2021xqp,Lenz:2020oce,Piscopo:2021ogu,Mannel:2020fts,Kirk:2017juj,Lenz:2013aua,Gabbiani:2004tp,Franco:2002fc,Beneke:2002rj,Lenz:2022rbq, 
Bag-parameters-new},
within the SM,
the most recent value of the ratio reads~\cite{NNLO}
\begin{equation}
   \left( \frac{\tau (B^+)}{\tau (B_d)}\right)^{\rm HQE}
    =
    1.081^{+0.014}_{-0.016} 
 \, ,
\label{eq:ratio_HQE}
\end{equation}
in perfect agreement with the experimental measurements, albeit with larger uncertainties.
In the presence of physics beyond the SM, new decay channels of the $b$-quark would also contribute to the total lifetime of the $B$
meson, and consequently modify the lifetime ratio according to 
 \begin{equation}
 \frac{\tau (B^+)}{\tau (B_d)}^{\rm HQE} \!\!\!\!\! = 1 + 
\left[\Gamma^{\rm SM} (B_d) - \Gamma^{\rm SM} (B^+) \right] 
\tau^{\rm Exp.} (B^+) 
+
\left[\Gamma^{\rm BSM} (B_d) - \Gamma^{\rm BSM} (B^+) \right]
\tau^{\rm Exp.} (B^+),
\label{eq:ratio_BSM}
\end{equation}
so that, by comparing with the corresponding experimental determination, Eq.~\eqref{eq:ratio_BSM} can be used to constrain the favoured parameter space for a specific set of BSM operators.
We determine the modification of the lifetime ratio due to the $B$-Mesogenesis model.
\\
The new decay channels also modify mixing observables like
the decay rate differences $\Delta \Gamma_q$ and the semileptonic CP asymmetries $a_{sl}^q$, with $q = d,s$. 
Currently SM predictions for the mixing observables
\cite{Davies:2019gnp,Dowdall:2019bea,FermilabLattice:2016ipl,DiLuzio:2019jyq,King:2019lal,Kirk:2017juj,Grozin:2016uqy,Lenz:2006hd,Beneke:2003az,Beneke:1998sy,Beneke:1996gn,Dighe:2001gc}
agree well with the experimental measurements, see e.g. HFLAV \cite{Amhis:2022mac}. The modification of the mixing observables due to the $B$-Mesogenesis model have been recently determined in Ref. \cite{Miro:2024fid} and turned out to be small. Nevertheless a future more precise determination of the  semileptonic CP asymmetries could have some impact on the $B$-Mesogenesis model.
\\
The paper is organised as follows: in Section \ref{sec:model} we briefly summarize the new $b$
quark decay channels within the $B$-Mesogenesis model. In Section \ref{sec:observables} we describe the new contributions to inclusive decay rates (Section \ref{subsec:inclusive}),
exclusive decay rates (Section \ref{subsec:exclusive})
and the lifetime ratios (Section \ref{subsec:exclusive}).
The numerical study of the effects of the $B$-Mesogenesis model is described in Section~\ref{sec:pheno}.
Finally, in
Section~\ref{sec:summary} we summarise our results and give an outlook on future developments.

\section{The model}
\label{sec:model}
\begin{figure}[ht]
    \centering
        \includegraphics[width=0.75\textwidth]{
        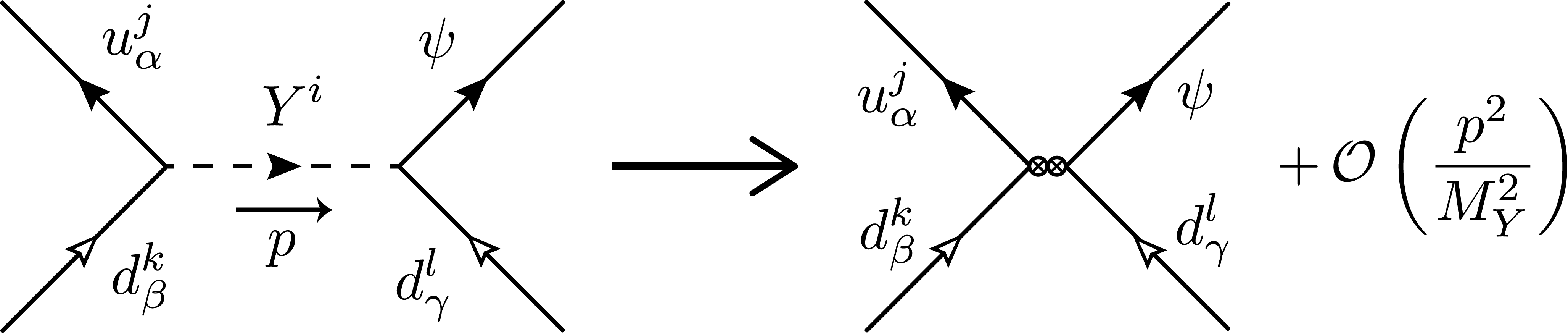}
    \caption{Diagramatic representation of integrating out the heavy scalar degree of freedom. Hollow arrows represent charge conjugated fermions, while filled-in arrows represent non-charge conjugated fermions.}
    \label{fig:HeavyScalar}
\end{figure}

\begin{figure}[ht]
    \centering
    \begin{subfigure}{0.45\textwidth}
        \centering
        \includegraphics[width=\textwidth]{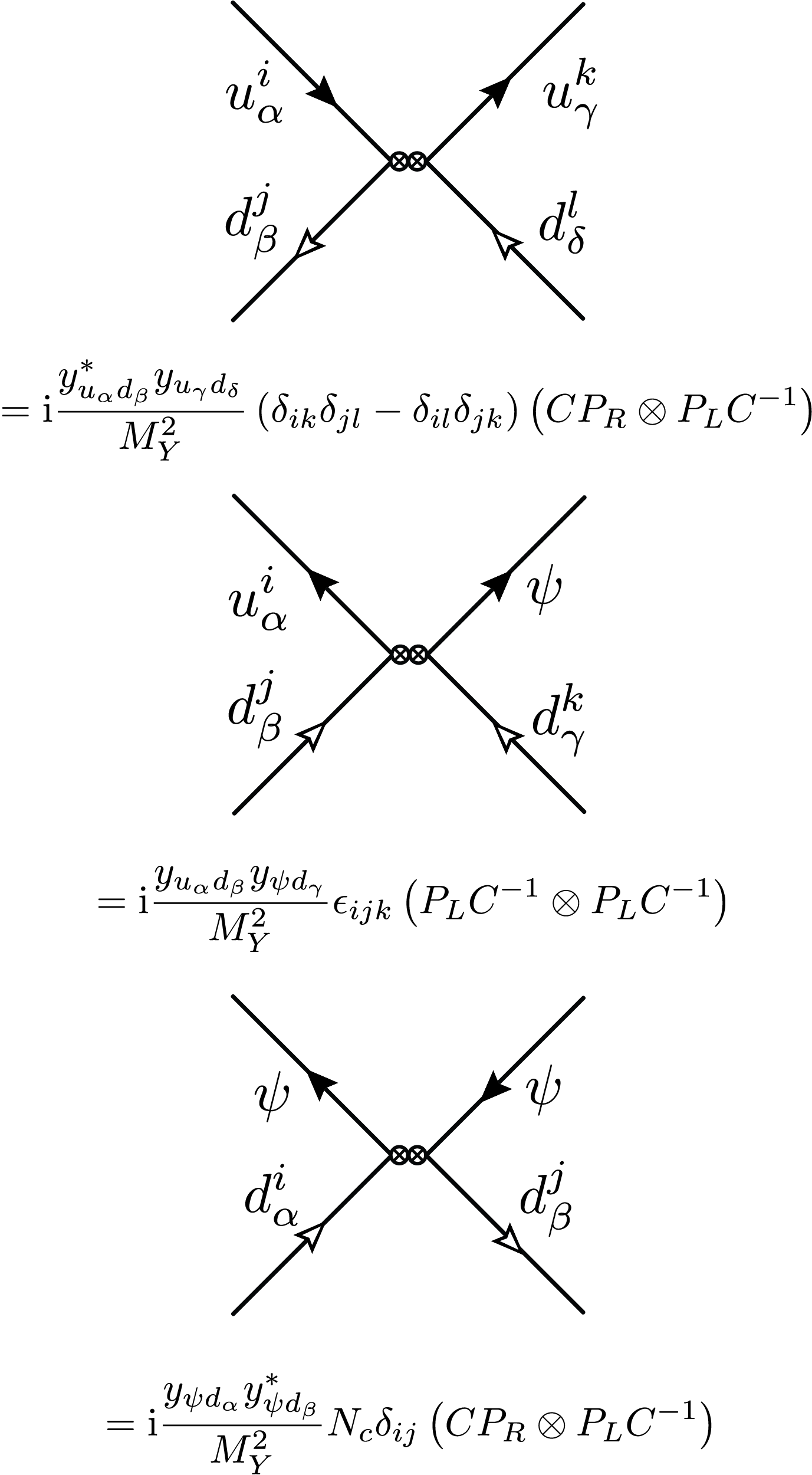}
        \caption{Effective vertices for $\mathcal{L}_{-1/3}$.}
    \end{subfigure}
    \hfill
    \begin{subfigure}{0.45\textwidth}
        \centering
        \includegraphics[width=\textwidth]{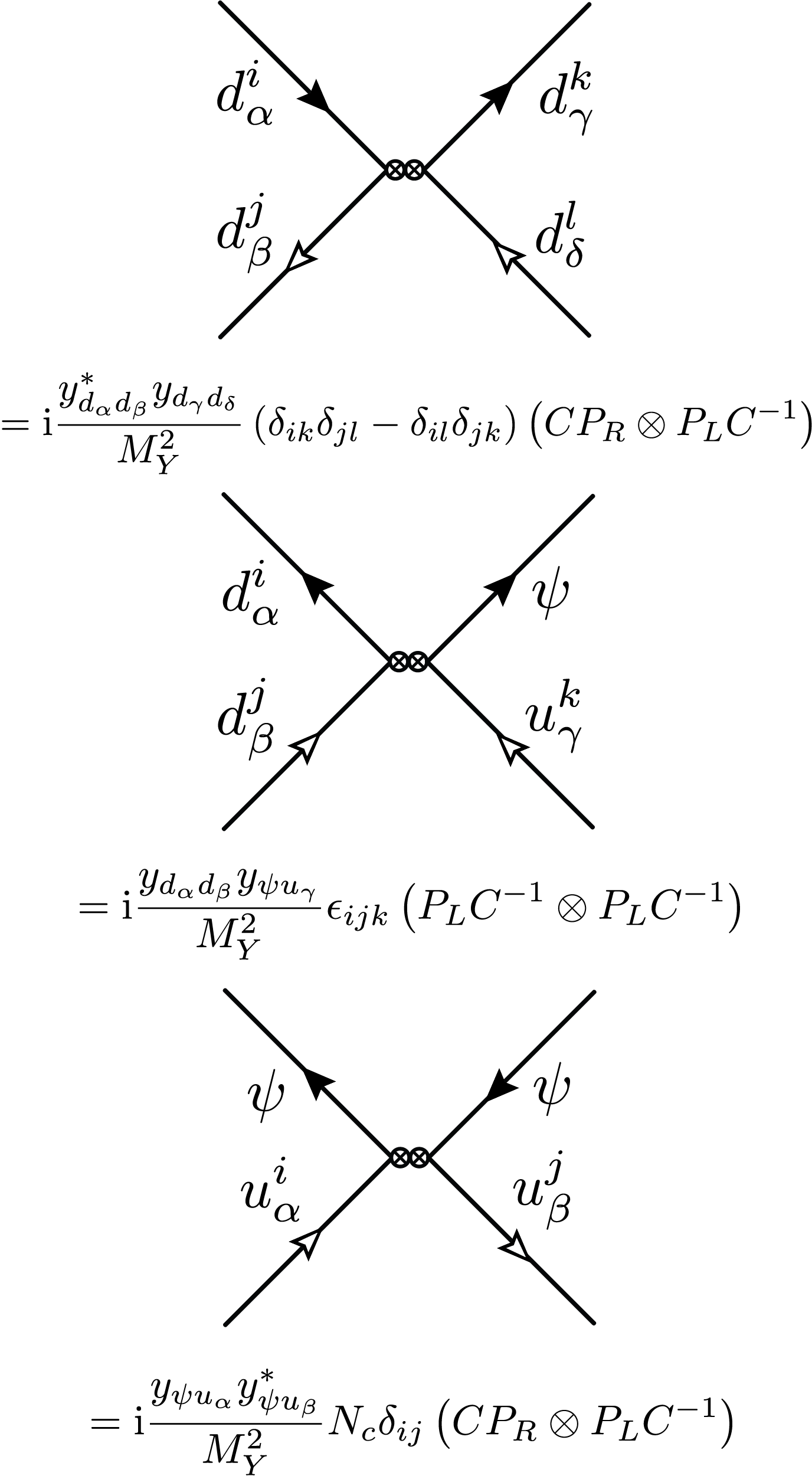}
        \caption{Effective vertices for $\mathcal{L}_{+2/3}$.}
    \end{subfigure}
    \caption{Possible effective vertices arising from integrating out the heavy scalar. Conjugate diagrams are not shown. As each effective vertex must be inserted in two fermion lines, the $(X \otimes Y)$ notation is meant to be interpreted as inserting $X$ in the left fermion line, and inserting $Y$ in the right fermion line (as seen in the diagram).}
    \label{fig:EffectiveVertices}
\end{figure}

The Baryogenesis model  \cite{Elor:2018twp,Alonso-Alvarez:2021qfd} is based on $B$ meson production in the early Universe, a process called "$B$-Mesogenesis". 
The starting point is the assumption that the early Universe is dominated by a combination of radiation and a very weakly coupled scalar particle $\Phi$, which will decay into 
$b\bar{b}$ pairs. 
The $b$-quarks then hadronize into $B$ and $\bar{B}$ meson pairs. In addition to standard decay channels, the $B$ and $\bar{B}$ mesons will subsequently decay through a new decay channel to visible baryons $\cal B$ and dark anti-baryons $\psi$, which will appear as missing energy.
Because of CP violation in the mixing of the neutral $B_d$ and $B_s$ systems, $B_q$ and $\bar{B}_q$ meson decays will result in different rates, 
generating the observed baryon asymmetry. 
Thus, this model not only directly relates the matter-antimatter imbalance to CP violation in the $B$ system, but it also suggests a dark matter candidate, all originating from the $B$ meson decays. 
\\
For the decay $ B \rightarrow \psi \mathcal{B}$ to occur, a heavy color-triplet scalar mediator $Y$ is introduced, which has baryon number $-2/3$ and can carry $-1/3$ or $+2/3$ hypercharge. The most general renormalizable Lagrangian describing the interaction of a (hypercharge $-1/3$ or $2/3$) color-triplet scalar with quarks and the SM singlet baryon $\psi$ is
\begin{equation}
\begin{aligned}
\mathcal{L}_{-1/3} = -\sum_{\alpha,\beta}  y_{u_{\alpha}d_{\beta}} \epsilon_{ijk} \kern0.18emY^{*i} \kern0.18em \bar{u}^{j}_{\alpha R}\kern0.18em d^{c,k}_{\beta R} - \sum_{\gamma} y_{\psi d_{\gamma}}  Y_{i} \kern0.18em \bar{\psi}\kern0.18em d^{c,i}_{\gamma R} + \text{h.c.}\kern0.18em,\\
\mathcal{L}_{+2/3} = -\sum_{\alpha,\beta}  y_{d_{\alpha}d_{\beta}} \epsilon_{ijk} \kern0.18emY^{*i} \kern0.18em \bar{d}^{j}_{\alpha R}\kern0.18em d^{c,k}_{\beta R} - \sum_{\gamma} y_{\psi u_{\gamma}}  Y_{i} \kern0.18em \bar{\psi}\kern0.18em u^{c,i}_{\gamma R} + \text{h.c.}\kern0.18em,
\label{eq:ModelLagrangians}
\end{aligned}
\end{equation}
where the $y$'s are the coupling constants, the subscripts $i,j,k$ are the color indices, $\alpha,\beta,\gamma$ denote the different flavours of the up and down quarks, and the superscript $c$ is the charge conjugate operator. We use the convention that $\psi^{c}_{R} = C^{-1}P_L^\top(\bar{\psi})^\top$ and $\bar{\psi}^{c}_{R} = - \psi^\top P_R^\top C$.
\\
The Lagrangian in Eq.~\ref{eq:ModelLagrangians} will be used below to determine new contributions to inclusive $B$ decays, to lifetimes of $B$ mesons, and to exclusive $B$ decays.  The first step in this process is to integrate out the new heavy scalar and express the new physics in terms of four fermion operators. This process is diagrammatically shown in Figure~\ref{fig:HeavyScalar}. Note that the charge conjugation operator can be moved between fermions on the same fermion line, represented diagrammatically by reversing arrow directions and swapping hollow and filled in arrows. Following this procedure of integrating out the heavy scalar, and considering every possible diagram at tree level with the heavy scalar, we arrive at the set of effective vertices given in Figure~\ref{fig:EffectiveVertices}.\footnote{Throughout the paper, we neglect QCD corrections in the derivation of the effective Hamiltonian, treating them as a higher-order approximation.}
\\
We can now easily reconstruct the necessary effective Hamiltonian to produce these effective vertices. In the case of the decay $B^+\to \psi p^+$ the relevant effective Hamiltonian is

\begin{equation}
   \mathcal{H}_{-1/3} =- \dfrac{y_{ub}y_{\psi d}}{M^{2}_{Y}} \kern0.14em \mathrm{i}\epsilon_{ijk}\bar{u}_{R}^{i} b^{c,j}_{R} \bar{d}^{k}_{R} \psi^{c}-\dfrac{y^{*}_{ub}y^{*}_{\psi d}}{M^{2}_{Y}} \kern0.14em \mathrm{i}\epsilon_{ijk}\bar{\psi}^{c} {d}_{R}^{i} \bar{b}^{c,j}_{R} {u}^{k}_{R} + \{ b \leftrightarrow d \},
\label{eq:HamiltonianInclusive13}
\end{equation}
which can be written as 
\begin{equation}
   \mathcal{H}_{-1/3} =- G_{(d)} \bar{\cal O}_{(d)}\kern0.14em\psi^{c}- G^{*}_{(d)} \bar{\psi}^{c}\kern0.14em {\cal O}_{(d)} + \{ b \leftrightarrow d \},
\label{eq:HamiltonianInclCompact}
\end{equation}
with the effective four-fermion coupling $G_{(d)} = \left({y_{ub}y_{\psi d}}\right)/{M^{2}_{Y}}$, and the local three-quark operators are defined as 
\begin{equation}
  \bar{\cal O}_{(d)} =\mathrm{i}\epsilon_{ijk} \left(\bar{u}_{R}^{i} b^{c,j}_{R}\right) \bar{d}^{k}_{R}, \quad {\cal O}_{(d)} =\mathrm{i}\epsilon_{ijk} {d}_{R}^{i} \left( \bar{b}^{c,j}_{R}  {u}^{k}_{R} \right).
\label{eq:Three-quark Operators(d)}
\end{equation}
It is also possible for the $b$-quark to couple to the dark matter particle $\psi$, which leads to
the following operators:
\begin{equation}
  \bar{\cal O}_{(b)} =\mathrm{i}\epsilon_{ijk} \left(\bar{u}_{R}^{i} d^{c,j}_{R}\right) \bar{b}^{k}_{R}, \quad {\cal O}_{(b)} =\mathrm{i}\epsilon_{ijk} {b}_{R}^{i} \left( \bar{d}^{c,j}_{R}  {u}^{k}_{R} \right).
\label{eq:Three-quark Operators(b)}
\end{equation}
For $\mathcal{L}_{2/3}$, the relevant part of the effective hamiltonian mediating the decay reads 
\begin{equation}
   \mathcal{H}_{2/3} =- G_{(u)} \kern0.14em \mathrm{i}\epsilon_{ijk}\bar{b}_{R}^{i} d^{c,j}_{R} \bar{\psi} u^{c,k}_{R}-\- G^*_{(u)} \kern0.14em \mathrm{i}\epsilon_{ijk}\bar{d}_{R}^{c,i} b^{j}_{R} \bar{u}_{R}^{c,k} \psi,
\label{eq:HamiltonianInclusive23}
\end{equation}
with the effective four-fermion coupling $G_{(u)} = \left({y_{db}y_{\psi u}}\right)/{M^{2}_{Y}}$.
\\
For determining the lifetime ratios in Section \ref{subsec:lifetime}, we will also include the four fermion operators involving two dark matter fields, or zero dark matter fields. However, these terms are irrelevant for the inclusive decay rate calculated in Section \ref{subsec:inclusive} and the exclusive decay rate calculated in Section \ref{subsec:exclusive}.

\begin{figure}[ht]
    \centering
    \begin{subfigure}{0.49\textwidth}
        \centering
        \includegraphics[width=\textwidth]{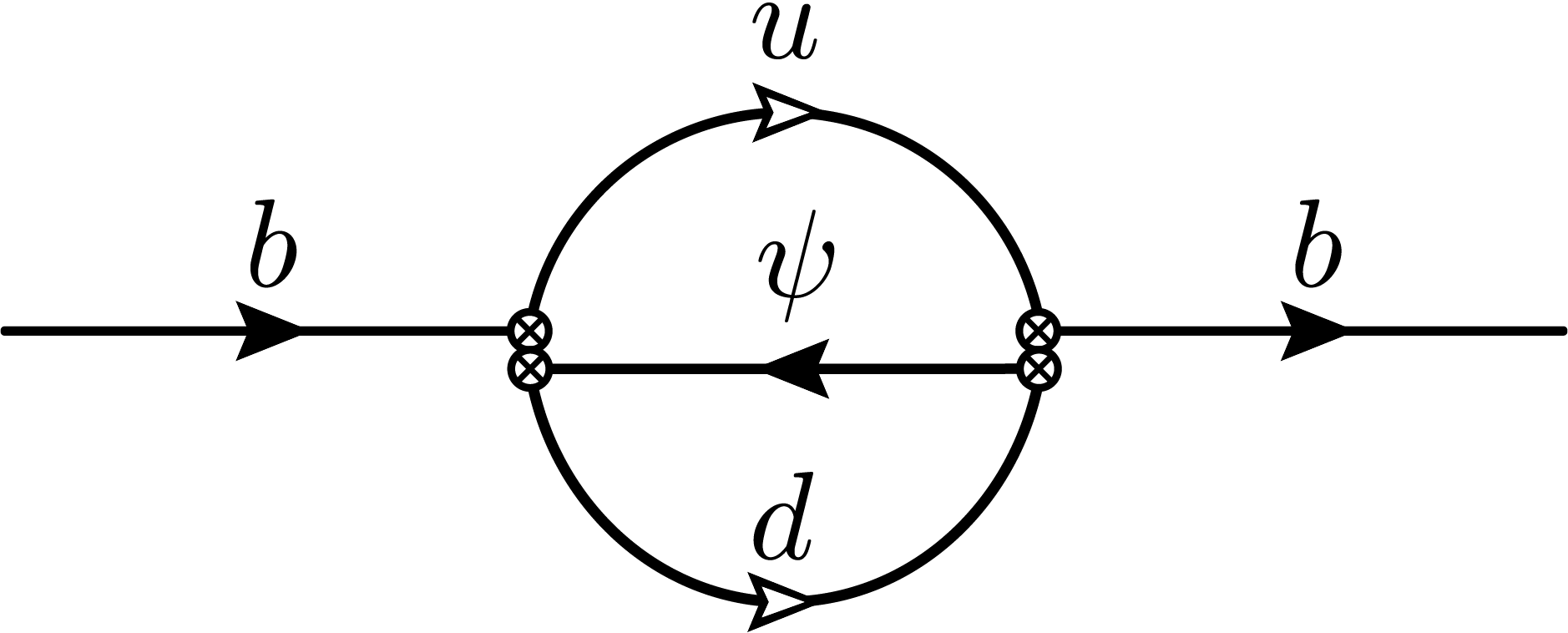}
    \end{subfigure}
    \hfill
    \begin{subfigure}{0.49\textwidth}
        \centering
        \includegraphics[width=\textwidth]{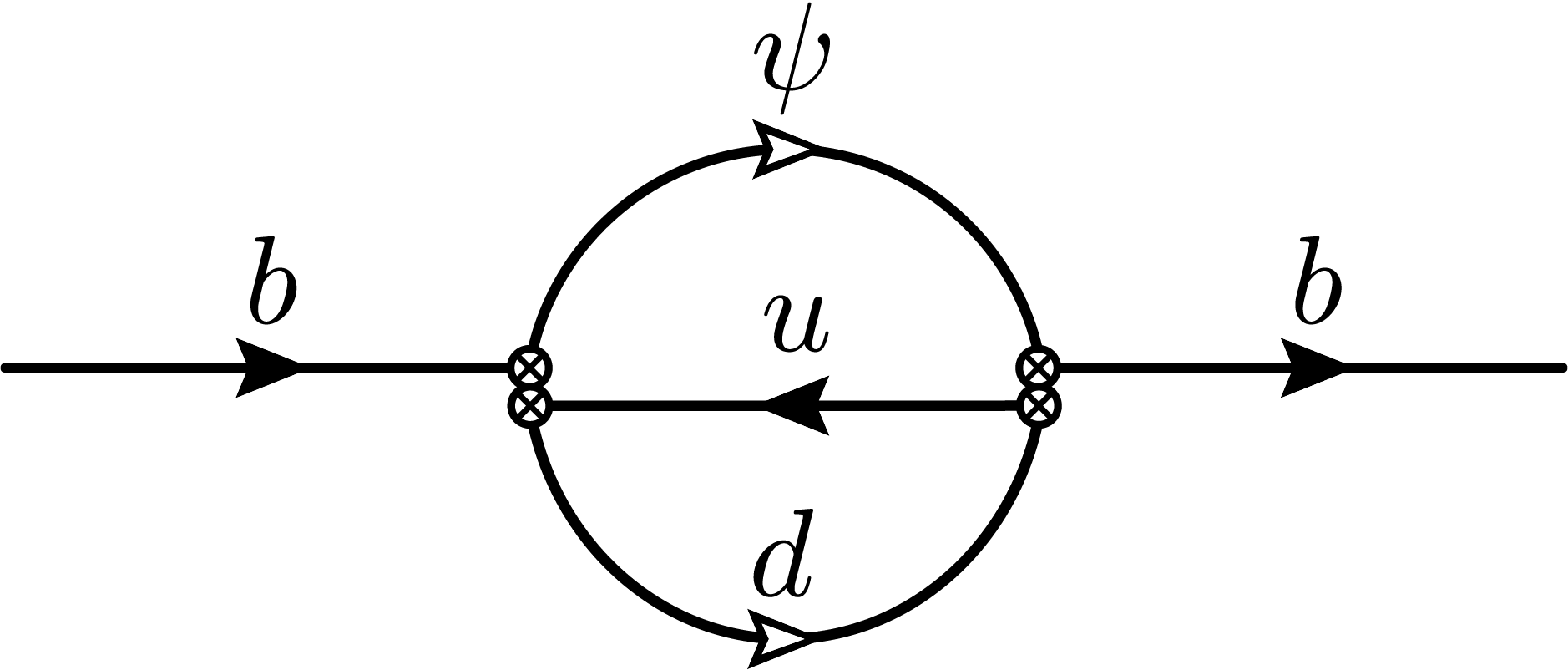}
    \end{subfigure}
    \caption{Diagrams contributing to the inclusive decay rate. Solid arrows indicate regular fermion lines, while hollow arrows indicate fermion lines involving the charge conjugation operator.}
    \label{fig:InclusiveRateDiagrams}
\end{figure}

\section{Contribution of the $B$-Mesogenesis model to flavour observables}
\label{sec:observables}
\subsection{Inclusive decay rate}
\label{subsec:inclusive}

In order to predict the right amount of matter and dark matter in the Universe, the $B$-Mesogenesis 
model has to provide a certain minimal strength of the new decay channels \cite{Alonso-Alvarez:2021qfd}. 
This requirement is quantitatively encoded in a lower limit of the inclusive branching ratio of the $B$ mesons into the dark baryon, i.e.
Br($B\to \psi  \cal B  \cal M $) $> 10^{-4}$, with a SM baryon $\cal B$, missing energy in the form of a dark sector antibaryon $\psi$, and
any number of light mesons denoted  by $\cal M$. 
\\
We calculate the  LO-QCD term of the dimension 3 decay rate $b \to d u \psi$ in the framework of HQE. The corresponding Feynman diagrams are shown in Figure~\ref{fig:InclusiveRateDiagrams}.
Our starting point is exactly the effective Hamiltonian $\mathcal{H}_{-1/3}$ in Eq.~\eqref{eq:HamiltonianInclCompact}, which mediates this decay. 
Following~\cite{Boushmelev:2023huu}, we consider the operators in Eqs.~\eqref{eq:Three-quark Operators(d)} and~\eqref{eq:Three-quark Operators(b)} as two individual
versions of the $B$-Mesogenesis model and call them $(d)$- and $(b)$-model, respectively. This corresponds to the ”type-2” and ”type-1” operators in~\cite{Alonso-Alvarez:2021qfd}, where it is shown that the flavour constraints on the heavy color-triplet scalar $Y$ imply that only one of the operators will be active and not a combination of both.\footnote{It should be noted that there is a third operator version referred to as ”type-3” in~\cite{Alonso-Alvarez:2021qfd}, where $\psi$ is coupled to the up-type quarks. However, this appears in $\mathcal{H}_{2/3}$.}
\\
By means of the optical theorem, the total decay width of a $B$ meson can be related to the imaginary part of the forward-scattering matrix element of the time-ordered product of the double insertion of the effective Hamiltonian, i.e.
\begin{equation}
    \Gamma (B) = \frac{1}{2 m_B}{\rm Im} \langle B|  {\cal T} | B \rangle \,,
    \label{eq:GammaB}
\end{equation}
with the transition operator given by
\begin{equation}
    {\cal T} = \mathrm{i} \int d^4 x \, T\, \{ {\cal H}_{\rm eff}(x), {\cal H}_{\rm eff}(0)\}\,.
    \label{eq:T-operator}
\end{equation}
The effective Hamiltonian ${\cal H}_{\rm eff}$ is given by the sum of the SM part and the $B$ mesogenesis part
\begin{eqnarray}
{\cal H}_{\rm eff} & = &
{\cal H}_{\rm eff}^{\rm SM} + 
{\cal H}_{-1/3} \, .
\end{eqnarray}
Within the HQE, the non-local operator in Eq.~\eqref{eq:T-operator} can be expressed as a systematic expansion in inverse powers of the heavy $b$-quark mass,
leading to the following series 
\begin{equation}
\Gamma(B) = 
\Gamma_3  +
\Gamma_5 \frac{\langle {\cal O}_5 \rangle}{m_b^2} + 
\Gamma_6 \frac{\langle {\cal O}_6 \rangle}{m_b^3} + \ldots  
 + 16 \pi^2 
\left( 
  \tilde{\Gamma}_6 \frac{\langle \tilde{\mathcal{O}}_6 \rangle}{m_b^3} 
+ \tilde{\Gamma}_7 \frac{\langle \tilde{\mathcal{O}}_7 \rangle}{m_b^4} + \ldots
\right)\,,
\label{eq:HQE}
\end{equation}
where $\Gamma_i$ are short-distance functions which can be computed perturbatively in QCD,
and $\langle {\cal O}_i \rangle \equiv
\langle B | {\cal O}_i |B  \rangle/(2 m_{B})$ denote the matrix element of local $\Delta B = 0$ operators of increasing dimension $i$. 
For now we are only interested in the leading term $\Gamma_3$, which describes the free $b$ quark decay. Below we will also consider higher order terms of Eq.~\eqref{eq:HQE} in order to investigate lifetime ratios of $B$ mesons.
\\
The LO-QCD contribution to the free quark decay of the channel $b\rightarrow du\psi$ reads
\begin{eqnarray}
 \Gamma^{(d)}_{3} (b \to d u \psi) & = &  \frac{|{G_{(d)}}|^2}{16} 
 \frac{  m_b^5}{192 \, \pi^{3}}
 \left[  1  - 8 \, \rho + 8\rho^3 - \rho^4 - 12 \, \rho^2 \, \text{log}(\rho)
 \right],
\label{eq:dim3 decayrateratio}
 \\
  \Gamma^{(b)}_{3} (b \to d u \psi) & = & 
   \Gamma^{(d)}_{3} (b \to d u \psi) 
 \Big|_{G_{(d)} \to \, G_{(b)}},
\nonumber
\end{eqnarray}
where $\rho = m^2_{\psi}/m^2_{b}$. We note, that this  result is identical to the muon decay if we make the replacements $|{G_{(d)}}|^2/16 \to G_F^2$, $m_b \to m_\mu$ and $m_{\psi} \to m_e$. We observe that as the mass of the dark particle increases, the available phase space for the decay products of the  $B$-meson becomes progressively smaller,
vanishing at  $m_{B^+} - m_p = 4341.14$ MeV. We expect the inclusive approach to hold at masses considerably lower than that bound. In our analysis the region below 3 GeV will be most relevant.
\\
For completeness, we also mention the result for $\mathcal{H}_{2/3}$ and we find the result of the inclusive decay to be equal to Eq.~\eqref{eq:dim3 decayrateratio} with the replacement 
\begin{eqnarray}
  \Gamma^{(2/3)}_{3} (b \to d u \psi) & = & 
   \Gamma^{(d)}_{3} (b \to d u \psi) 
 \Big|_{G_{(d)} \to \, G_{(u)}}.
 \label{eq:dim3 decayrateratio2/3}
\end{eqnarray}

\subsection{Exclusive decay rates}
\label{subsec:exclusive}
The exclusive decay for the channel $B \to p^{+} \psi$ was calculated for  $\mathcal{H}_{-1/3}$ in Ref.~\cite{Boushmelev:2023huu} using QCD light-cone sum rules up to twist six\footnote{An exclusive decay analysis based  on $\mathcal{H}_{2/3}$ has not yet been performed.}. The exclusive decay width reads 
\begin{align}
	&\Gamma_{(d)}(B^+\to p\psi) =
	|G_{(d)}|^2 \Bigg\{\Bigg[\Big(F^{(d)}_{B\to p_R}(m_\psi^2)\Big)^2
	+ \frac{m_\psi^2}{m_p^2}\Big(\widetilde{F}^{(d)}_{B\to p_L}(m_\psi^2)\Big)^2\Bigg] \nonumber \\
	&\times \big(m_B^2-m_p^2-m_\psi^2\big)
	+ \;
	2 m_\psi^2F^{(d)}_{B\to p_R}(m_\psi^2)\widetilde{F}^{(d)}_{B\to p_L}(m_\psi^2)\Bigg\}
	\,\frac{\lambda^{1/2}(m_B^2,m_p^2,m_\psi^2)}{16\pi m_B^3}\, ,
	\label{eq:widthd}
\end{align}\noindent
where $F^{(d)}_{B \to p_R}(q^2)$ and $ \widetilde{F}^{(d)}_{B\to p_L} (q^2)$ are the form factors
\begin{align}
	F^{(d)}_{B \to p_R}(q^2) =& \; \frac{F^{(d)}_{B \to p_R}(0)}{1 - q^2/m_{\Lambda_b}^2} \Bigg[1 + b^{(d)}_{B \to p_R} \bigg(z(q^2) - z(0) + \frac{1}{2} \Big[z(q^2)^2 - z(0)^2\Big]\bigg)\Bigg]\,,  \label{eq:FFzExpFinal}
\end{align}
and $ \widetilde{F}^{(d)}_{B\to p_L} (q^2)$ is obtained by the replacement $F^{(d)}_{B \to p_R}(q^2) \to \, \widetilde{F}^{(d)}_{B\to p_L} (q^2)$. Moreover 
\begin{align}
    z(q^2)=(\sqrt{t_+-q^2}-\sqrt{t_+-t_0})/(\sqrt{t_+-q^2}+\sqrt{t_+-t_0}),
\end{align}
with $t_0 = (m_B+m_p)\cdot (\sqrt{m_B}-\sqrt{m_p})^2$ and $t_\pm=(m_B\pm m_p)^2$, where $m_{B}$, $m_{p}$ and $m_{\Lambda_{b}}$ are the masses of the $B$ meson, the proton and the 
$\Lambda_{b}$ baryon, respectively. $\lambda(x,y,z) = x^2+y^2+z^2-2xy-2xz-2yz$ is the Källen function, and $b^{(d)}_{B \to p_R}$ is the slope parameter obtained from the fitting procedures, for more details see~\cite{Boushmelev:2023huu}. The corresponding relation for the model (b) is obtained by a simple exchange of (d) with (b). The numerical values of the various parameters are listed in Table \ref{tab:input}.
\begin{table}[h]
	\centering
	\scalebox{0.95}{
	\begin{tabular}{|c||c|c|}
		\hline
		&&\\[-3.0mm]
		Parameter  &  interval & Ref.  \\
		&&\\[-3.0mm]
		\hline
		\hline
		&&\\[-3.0mm]
		$b$-quark $\overline{\text{MS}}$ mass & $\overline{m}_b(3~\mbox{GeV})= 4.47^{+0.04}_{-0.03} $ GeV 
		& \cite{ParticleDataGroup:2020ssz}
		\\
		&&\\[-3.0mm]
		\hline
		&&\\[-3.0mm]
		 &$F^{(d)}_{B\to p_R}(0) = 0.022^{+0.013}_{-0.013}$ $\text{GeV}^2$&
		\multirow{3.25}{*}{\cite{Boushmelev:2023huu}}
		\\[1mm]
		Form Factors  &  $\widetilde{F}^{(d)}_{B\to p_L}(0) = 0.005_{-0.001}^{+0.002}\, \text{GeV}^2$ &\\[1mm]
		&  $F^{(b)}_{B\to p_R}(0) = -0.041_{-0.018}^{+0.019}$  GeV$^2$ &
		 \\[1mm]
		 &  $\widetilde{F}^{(b)}_{B\to p_L}(0) = -0.007_{-0.002}^{+0.003}\, \text{GeV}^2$ & \\
		&&\\[-3.0mm]
		\hline
		&&\\[-3.0mm]
		 & $b^{(d)}_{B\to p_R} = 4.46^{+0.97}_{-1.72} \, \text{GeV}^2$ & 
		\multirow{3.25}{*}{\cite{Boushmelev:2023huu}}
		\\[1mm]
		 Slope parameters &  $b^{(d)}_{B\to p_L} = -2.27_{-0.08}^{+0.10}\, \text{GeV}^2$ &
		 \\[1mm]
		 &  $b^{(b)}_{B\to p_R}  = -2.00_{-3.62}^{+1.58}\, \text{GeV}^2$ &
		 \\[1mm]
		 &  $b^{(b)}_{B\to p_L}  = -2.85_{-0.15}^{+0.17}\, \text{GeV}^2$ &
		\\
		\hline
	\end{tabular}}
	\caption{Input parameters in the LCSRs from the references in this Table.}
	\label{tab:input}
\end{table} \noindent
We see that these decay rates are also proportional to $|{G_{(d)}}|^2$ or $|{G_{(b)}}|^2$.
\\
Therefore we start with the necessary bound on the inclusive branching ratio for the $B$ mesogenesis model to work
\cite{Alonso-Alvarez:2021qfd}
\begin{eqnarray}
Br(B^+\to \psi  \cal B  \cal M ) 
& > & 10^{-4}\, ,
\nonumber
\\
 Br( B^{+} \to p^{+} \psi)
& > & 10^{-4} \frac{\Gamma(B^{+} \to p^{+} \psi)}{\Gamma(B^+\to \psi  \cal B  \cal M )}
\equiv 10^{-4} \frac{\Gamma(B^{+} \to p^{+} \psi)}{\Gamma (b \to d u \psi)}
\label{eq:ratio_decayrates}
\end{eqnarray}
to get a lower bound on the  decay channel $B^{+} \to p^{+} \psi$.
In the ratio of the decay rates in the last line of Eq. (\ref{eq:ratio_decayrates}) the unknown couplings  $|{G_{(d)}}|^2$ or $|{G_{(b)}}|^2$ cancel.

\subsection{Lifetime differences}
\label{subsec:lifetime}

\begin{figure}[ht]
    \centering
        \centering
        \includegraphics[width=\textwidth]{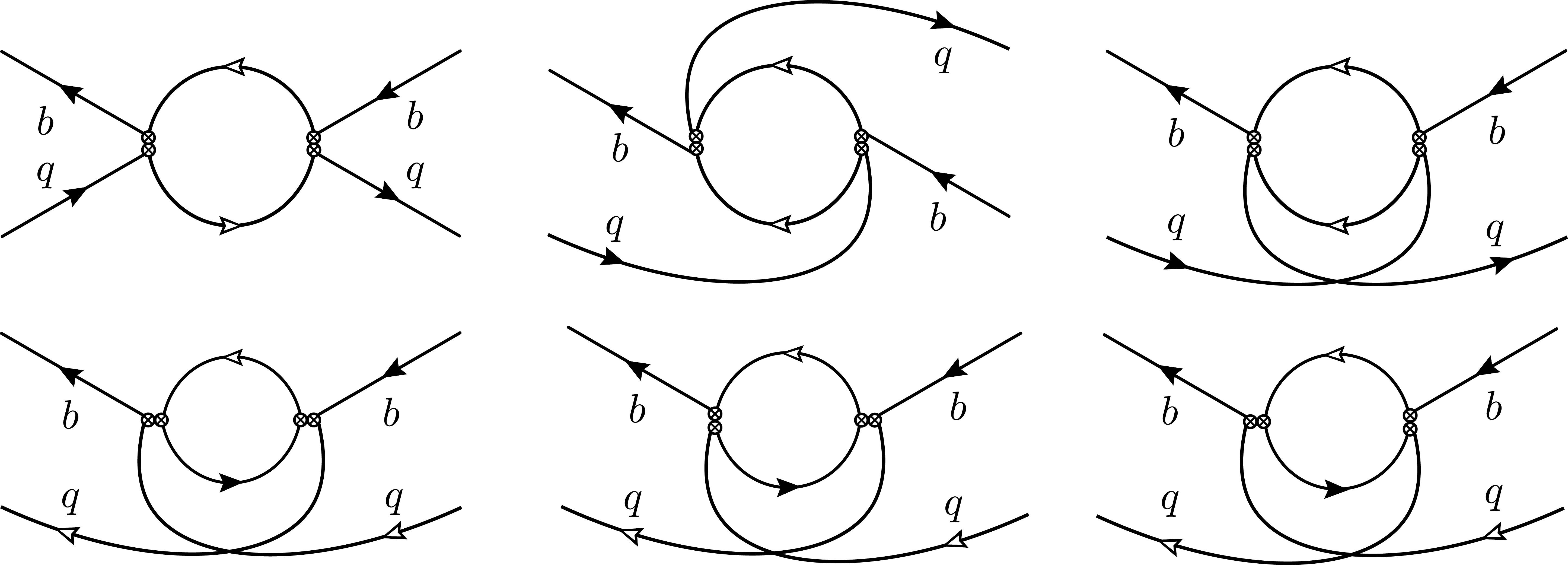}
        \caption{All the possible topologies that can show up in Eq.~\eqref{eq:T-NP}. From left to right, the top row corresponds to topologies $I$, $II$, and $III$ respectively and the bottom row corresponds to topologies $IV$, $V$, and $VI$ respectively. The possible internal fermions in the loop for each diagram depend on the specific effective vertices used. The list of possible effective vertices is provided in Figure \ref{fig:EffectiveVertices}.}
    \label{fig:Topologies}
\end{figure}

Finally we study the implications of the new decay channels of the $b$-quarks to the lifetime ratio $ {\tau (B^+)}/{\tau (B_d)}$.
For this quantity all 2-quark contributions
in Eq.~\eqref{eq:HQE}, i.e. the terms proportional to  $\langle {\cal O}_5 \rangle$, are cancelling in the differences  in Eq.~\eqref{eq:ratio_BSM} due to isospin symmetry. Thus in   Eq.~\eqref{eq:ratio_BSM} only four-quark operator contributions survive  starting from dimension-six,  originating from loop-enhanced diagrams, as reflected by the explicit factor of $16 \pi^2$ in Eq.~\eqref{eq:HQE}.\footnote{In fact, this is not the case for the lifetime ratio $\tau(B_s)/\tau(B_d)$, where SU(3)$_F$ breaking effects play a dominant role. Therefore, a detailed study of BSM contributions to  $\tau(B_s)/\tau(B_d)$ would currently be strongly limited by the size of the non-perturbative input which parametrise the matrix elements of two-quark operators, and even more of the corresponding SU(3)$_F$ breaking effects, which are poorly known, particularly for the Darwin operator, see e.g.\ the recent work~\cite{Lenz:2022rbq}. Hence, given the current status of the SM prediction, we postpone the study of NP effects in $\tau(B_s)/\tau(B_d)$ to a future work, once further insights on the size of matrix element of the Darwin operator and 
of the SU(3)$_F$ breaking effects will become available. } More details on the structure of the HQE for the $b$-system, as well as a complete list of references can be found e.g.\ in Ref.~\cite{Lenz:2022rbq}.
\\
The leading (dimension-six and LO-QCD)  contribution of the new decay channels in 
 the imaginary part of the transition operator in Eq.~\eqref{eq:T-operator} can be compactly written as
\begin{equation}
{\rm Im}{\cal T}_{X,\rm{NP}}^{f_1f_2} = \frac{ 2 m_b^2 \, G^{f_1f_2}_{X} \sqrt{\lambda_{f_1 f_2}}}{192 \pi}\,
\left[\,
\sum_{n = 1}^{4} \, A^{f_1 f_2}_{X,n}\, \tilde{O}_{n}
\,
\right].
\label{eq:T-NP}
\end{equation}
Here $X$ labels the topology (see Figure~\ref{fig:Topologies}), while $f_1$ and $f_2,$ are the internal particles running in the loop, i.e. the light quarks or the dark matter particle $\psi$.
The constants $G^{f_1f_2}_{X}$ are defined in Tables \ref{Constants1/3} and \ref{Constants2/3} in Appendix \ref{app:A} for each topology and $\lambda_{f_1f_2} = (1-r_{f_1}-r_{f_2})^{2}-4r_{f_1}r_{f_2}$, with $r_{f}={m^{2}_{f}}/{m^{2}_{b}}$. 
We also define the parameter $\tilde{\lambda}_{f_1f_2} \equiv 2(r_{f_1}-r_{f_2})^{2}-(1+r_{f_1}+r_{f_2})$. 
The functions~$A^{f_1 f_2}_{X,n}$ in Eq.~\eqref{eq:T-NP} are then linear combinations of the parameters $\lambda_{f_1 f_2}$ and $\tilde{\lambda}_{f_1f_2}$.
The $\Delta B = 0$ four-quark operators $\tilde{O}_{n}$ are defined as \begin{align}
& \tilde{Q}_1 = 
( \bar{q}^i \gamma_\mu P_R \, b^i) 
(\bar b^j \gamma^\mu P_R \, q^j)\,,
\quad &
& \tilde{Q}_3 = 
(\bar q^i \gamma_\mu P_R t^{a}_{ij} \, b^j) 
(\bar b^r \gamma^\mu P_R t^{a}_{rm} \, q^m)\,,
\label{eq:Q1-Q2}
\\[2mm]
& \tilde{Q}_2 = 
(\bar q^i P_L \, b^i) 
(\bar b^j P_R \, q^j)\,,
\quad &
& \tilde{Q}_4 = 
(\bar q^i P_L t^{a}_{ij}\, b^j) 
(\bar b^r P_R t^{a}_{rm} \, q^m)\,. 
\label{eq:Q3-Q4}
\end{align}  
where $q$ is $u (d)$ for $\Gamma(B_d) \, (\Gamma(B^+))$. It is worth noting that due to the fact that the values of the constants $G^{f_1f_2}_{X}$ corresponding to topologies where $f_1,f_2 \neq \psi$ are extremely suppressed, the contributions of diagrams with both a BSM vertex and a SM vertex are negligible. 

\subsubsection{Results for ${\cal H}_{-1/3}$ }
For $\Gamma(B_d)$, the analytic expressions for the entries $ A^{f_1 f_2}_{X,n}$ in Eq.~\eqref{eq:T-NP} read
\begin{eqnarray}
& \displaystyle
A_{III,1}^{\psi c}  = \frac{2}{3} \, (-\lambda_{\psi c}+\tilde{\lambda}_{\psi c})  , \quad  
A_{III,2}^{\psi c}  = \frac{-4}{3} \, \tilde{\lambda}_{\psi c} ,  \quad
A_{III,3}^{\psi c} = -3A_{III,1}^{\psi c}, \quad
A_{III,4}^{\psi c} = -3A_{III,2}^{\psi c},
\end{eqnarray}
\begin{eqnarray}
& \displaystyle
A_{II,1}^{\psi c}  =  \, \frac{2}{3}  \,\lambda_{\psi c}, \quad  
A_{II,2}^{\psi c}  = \frac{-4}{3} \, \tilde{\lambda}_{\psi c} ,  \quad
A_{II,3}^{\psi c} = -3A_{II,1}^{\psi c}, \quad
A_{II,4}^{\psi c} = -3A_{II,2}^{\psi c},
\end{eqnarray}
\begin{eqnarray}
& \displaystyle
A_{I,1}^{\psi \psi}  = - \, (\lambda_{\psi \psi}-\tilde{\lambda}_{\psi \psi})  , \quad  
A_{I,2}^{\psi \psi}  = -2 \, \tilde{\lambda}_{\psi \psi} ,  \quad
A_{I,3}^{\psi \psi} = A_{I,4}^{\psi \psi} = 0,
\end{eqnarray}
\begin{eqnarray}
& \displaystyle
A_{I,1}^{c c}  = -\frac{4}{3} \, (
\lambda_{c c}-\tilde{\lambda}_{c c})  , \quad  
A_{I,2}^{c c}  = -\frac{8}{3} \, \tilde{\lambda}_{c c} ,  \quad A_{I,3}^{c c} = \frac{3}{2}A_{I,1}^{c c},\quad A_{I,4}^{c c} = \frac{3}{2}A_{I,2}^{c c},
\end{eqnarray}
while the remaining coefficient functions are obtained as follows
\begin{equation}
\begin{aligned}
A_{X,n}^{f_1 f_2} = A_{X,n}^{f_2 f_1},\qquad A_{X,n}^{f_1 f_3}  = A_{X,n}^{f_1 f_2}\Big|_{\lambda_{f_1 f_2},\tilde{\lambda}_{f_1 f_2},r_{f_2} \to {\lambda_{f_1 f_3},\tilde{\lambda}_{f_1 f_3},r_{f_3}}}
\end{aligned}
\end{equation}
In the case of $\Gamma(B^+)$, the corresponding analytic expressions for $ A^{f_1 f_2}_{X,n}$ in Eq.~\eqref{eq:T-NP} read
\begin{eqnarray}
& \displaystyle
A_{IV,1}^{s \psi}  = \frac{-8}{3} \, \lambda_{s \psi }+ \frac{4}{3}\tilde{\lambda}_{s \psi}  , \quad  
A_{IV,2}^{s \psi }  = 0 ,  \quad
A_{IV,3}^{s \psi} = -3A_{IV,1}^{s \psi}, \quad
A_{IV,4}^{s \psi } = 0,
\end{eqnarray}
\begin{eqnarray}
& \displaystyle
A_{III,1}^{ \psi s}  = \frac{-2}{3}  \,\lambda_{\psi s}+\frac{2}{3}\tilde{\lambda}_{\psi s}  , \quad
A_{III,2}^{\psi s }  = \frac{-4}{3}  \tilde{\lambda}_{\psi s} ,  \quad 
A_{III,3}^{\psi s} = -3A_{III,1}^{\psi s}, \quad
A_{III,4}^{\psi s} = -3A_{III,2}^{\psi s},
\end{eqnarray}
\begin{eqnarray}
& \displaystyle
A_{V,1}^{s \psi}  = \frac{8}{6} \, \lambda_{s \psi}- \frac{4}{6}\tilde{\lambda}_{s \psi}  , \quad  
A_{V,2}^{s \psi}  = 0,  \quad
A_{V,3}^{s \psi} = \frac{-6}{2}A_{V,1}^{s \psi} , \quad  
A_{V,4}^{s \psi} = 0,
\end{eqnarray}
\begin{eqnarray}
& \displaystyle
A_{VI,n}^{s \psi}  = A_{V,n}^{s \psi}, 
\end{eqnarray}
\begin{eqnarray}
& \displaystyle
A_{IV,1}^{s c}  = -\frac{2}{3} \, (8
\lambda_{s c}-4\tilde{\lambda}_{s c})  , \quad  
A_{IV,2}^{s c}  = 0 ,  \quad
A_{IV,3}^{s c} = -3A_{IV,1}^{s c},\quad
A_{IV,4}^{s c} = 0,
\end{eqnarray}
\begin{eqnarray}
& \displaystyle
A_{I,1}^{s c}  = -\frac{8}{6} \, (
\lambda_{s c}-\tilde{\lambda}_{s c})  , \quad  
A_{I,2}^{s c}  = -\frac{8}{3} \, \tilde{\lambda}_{s c},  \quad
A_{I,3}^{s c} = \frac{3}{2}A_{I,1}^{s c},\quad
A_{I,4}^{s c} = \frac{3}{2}A_{I,2}^{s c},
\end{eqnarray}
The remaining functions are obtained from the following replacement
\begin{equation}
\begin{aligned}
A_{X,n}^{f_1 f_3}  = A_{X,n}^{f_1 f_2}\Big|_{\lambda_{f_1 f_2},\tilde{\lambda}_{f_1 f_2},r_{f_2} \to {\lambda_{f_1 f_3},\tilde{\lambda}_{f_1 f_3},r_{f_3}}}.
\end{aligned}
\end{equation}
\subsubsection{Results for ${\cal H}_{2/3}$ }
For $\Gamma(B_d)$, the analytic expressions for the entries $ A^{f_1 f_2}_{X,n}$ in Eq.~\eqref{eq:T-NP} read
\begin{eqnarray}
& \displaystyle
A_{IV,1}^{\psi c}  = \frac{2}{3} \, (-4\lambda_{\psi c}+2\tilde{\lambda}_{\psi c}),\quad
A_{IV,3}^{\psi c} = -3A_{IV,1}^{\psi c}, \quad  
A_{IV,2}^{\psi c}  = A_{IV,4}^{\psi c} = 0,
\end{eqnarray}
\begin{eqnarray}
& \displaystyle
A_{IV,1}^{d s}  = \frac{-2}{3} \, (8\lambda_{d s}-4\tilde{\lambda}_{d s}),\quad
A_{IV,3}^{d s} = -3A_{IV,1}^{d s}, \quad  
A_{IV,2}^{d s}  = A_{IV,4}^{d s} = 0,
\end{eqnarray}
\begin{eqnarray}
& \displaystyle
A_{I,1}^{s s}  = \frac{-8}{6} \, (\lambda_{s s}+\tilde{\lambda}_{s s}),\quad
A_{I,2}^{s s} = \frac{16}{6} \tilde{\lambda}_{s s}, \quad  
A_{I,3}^{s s}  = \frac{3}{2} A_{I,1}^{s s},\quad  
A_{I,4}^{s s}  = \frac{3}{2} A_{I,2}^{s s},
\end{eqnarray}
while the remaining coefficient functions are obtained as follows
\begin{equation}
\begin{aligned}
A_{IV,n}^{\psi u}  = A_{IV,n}^{\psi c}\Big|_{\lambda_{\psi c},\tilde{\lambda}_{\psi c},r_{c} \to {\lambda_{\psi u},\tilde{\lambda}_{\psi u},r_{u}}},\quad   A_{I,n}^{s d}  = A_{I,n}^{s s}\Big|_{\lambda_{s s},\tilde{\lambda}_{s s},r_{s} \to {\lambda_{s d},\tilde{\lambda}_{s d},r_{d}}}.
\end{aligned}
\end{equation}
In the case of $\Gamma(B^+)$, the corresponding analytic expressions for $ A^{f_1 f_2}_{X,n}$ in Eq.~\eqref{eq:T-NP} read
\begin{eqnarray}
& \displaystyle
A_{III,1}^{s \psi}  = \frac{-2}{3} \, (\lambda_{s \psi}-\tilde{\lambda}_{s \psi }),\quad
A_{III,2}^{s \psi} = \frac{-4}{3}\tilde{\lambda}_{s \psi } , \quad  
A_{III,3}^{ s \psi}  = -3 A_{III,1}^{s \psi},\quad  
A_{III,4}^{ s \psi}  = -3 A_{III,2}^{s \psi}.
\end{eqnarray}
The remaining functions are obtained from the following replacement
\begin{equation}
\begin{aligned}
A_{III,n}^{d \psi}  = A_{III,n}^{s \psi}\Big|_{\lambda_{s \psi},\tilde{\lambda}_{s \psi},r_{s} \to {\lambda_{d \psi},\tilde{\lambda}_{d \psi},r_{d}}}.
\end{aligned}
\end{equation}

\subsubsection{Matrix elements of four quark operators}

We discuss now the parametrisation of the matrix element of the $\Delta B = 0$ four-quark operators.
The new operators arising in 
Eq.~\eqref{eq:Q1-Q2} and 
Eq.~\eqref{eq:Q3-Q4}  differ from the SM ones only by a chirality transformation and therefore their matrix elements yield the same results as in the SM.
We stress that in order to be consistent with the SM prediction of $\tau(B^+)/\tau(B_d)$, obtained in Ref.~\cite{NNLO}, and which we use in our analysis, we also parametrise the operators in HQET. In fact, any difference between operators defined in QCD or HQET arises only at dimension-seven, which we do not include in the present work. We thus have
\begin{equation}
\langle B | \tilde{\Oq}_n | B \rangle 
= f_B^2 \, m_B^2 \, \B_n \, ,
\label{eq:Par-ME}
\end{equation} 
where $f_B$ is the QCD decay constant, 
and $\B_n$ denote the corresponding Bag parameters. 
Within vacuum insertion approximation (VIA), it is easy to show that 
\begin{equation}
\B_1 = \B_2 = 1\,, \qquad 
\B_3 = \B_4  = 0\,.
\label{eq:Bag-VIA}
\end{equation}
We emphasise that for the Bag parameters $\B_{i}\,,$ with $i= 1, \ldots, 4$, also computations based on HQET sum rule are available~\cite{Kirk:2017juj, King:2021jsq,Bag-parameters-new}, however, the deviation from their VIA values is found to be small, at most of the order of few percents. In our numerical analysis, we use for~$\B_{i}\,,$ with $i= 1, \ldots, 4$, the determination from Ref.~\cite{King:2021jsq}.

\section{Phenomenology}
\label{sec:pheno}

\subsection{Inclusive vs. exclusive decays}
\begin{figure}[ht]
    \centering
    \begin{subfigure}{0.49\textwidth}
        \centering
        \includegraphics[width=\textwidth]{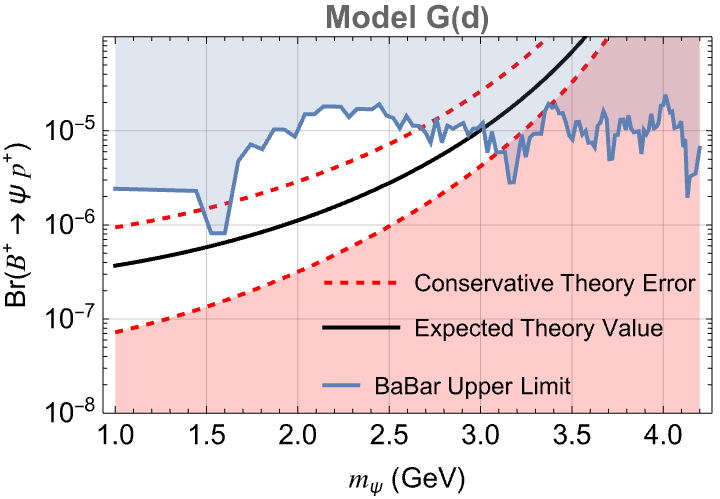}
    \end{subfigure}
    \hfill
    \begin{subfigure}{0.49\textwidth}
        \centering
        \includegraphics[width=\textwidth]{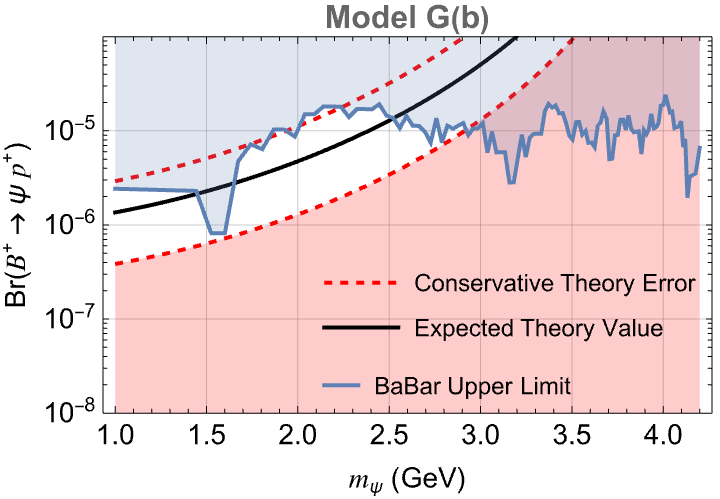}
    \end{subfigure}
    \caption{ The red shaded region corresponds to the excluded region if one assumes the $B$ mesogensis model to work. The blue shaded region corresponds to the $90\%$ CL excluded region from the BABAR upper limit of $\text{Br}(B^+ \to p^{+} \psi)$. The remaining allowed region is given in white.}
    \label{fig:ExclusionPlots}
\end{figure}

 First we study the lower limit on the exclusive decay $\Gamma_{\text{exl}}(B \to p^{+} \psi)$ obtained in Eq.~\eqref{eq:ratio_decayrates}
 as a function of the mass of the dark matter particle $m_{\psi}$ in Figure~\ref{fig:ExclusionPlots}. 
 Note that the unknown couplings $G_{(d,b)}$ are cancelling.
 In estimating the theory uncertainties, we follow two approaches. 
 The first approach is a conservative scenario where we consider the maximal deviation of the prediction from the central values, when varying all input parameter
 in the region given in Table
 \ref{tab:input}.
 The second approach is adding the uncertainties in quadrature, and this results in similar overall uncertainties. 
 \\
 In an attempt to search for direct evidence of the
 $B$-Mesogenesis model,
 the BABAR Collaboration conducted a study of the
 decay $B^+ \to \psi p^+$~\cite{BaBar:2023dtq}.
 Although no direct signal was observed, 
 they established an upper bound on the branching ratio $\mathrm{Br}(B^+ \to \psi p^+)$, 
 as illustrated in 
 Figure~\ref{fig:ExclusionPlots}.
\\
For the $B$-Mesogensis model to be valid the
branching fraction of the $B^+ \to p^+ \psi$ decay
has to be above the red dotted line. 
Experimentally branching fractions of this decay
inside the blue area have already been excluded.
The the remaining allowed region is given in white
and indicates that masses for the dark matter particle $\psi$ above 3 GeV are excluded.
For masses below 3 GeV there is still some possibility for the  $B$-Mesogenesis model to work, which can, however, be tested by future, more precise experimental studies.

\subsection{Lifetime ratios}
Next we investigate the impact of the 
$B$-Mesogenesis model on the lifetime ratio 
${\tau (B^+)}/{\tau (B_d)}$. 
To constrain the numerical values of the parameters
of the $B$-Mesogenesis model, we note that the
studies  in~\cite{Alonso-Alvarez:2021qfd} have
shown that ATLAS and CMS searches for color-triplet
scalar impose stringent constraints on the
constants $G^{f_1f_2}_{X}$. 
We use the maximum values of these constants listed
in Tables \ref{Constants1/3} and \ref{Constants2/3}
to plot the lifetime ratio as a function of the mass of the dark sector particle $\psi$,
 as shown in Figures~\ref{fig:Lifetime1/3-BdAllPlots},~\ref{fig:Lifetime1/3-BPlusAllPlots} and~\ref{fig:Lifetime2/3AllPlots} in Appendix \ref{supplement}.
 There we have turned on one constant at a time, while all others are turned off.
 Unfortunately we find that within uncertainties the theory prediction
 overlaps always with the experimental value for all values of $m_\psi$. 
 Hence we conclude that the study of the implications of the $B$-Mesogenesis model on the lifetime ratio ${\tau (B^+)}/{\tau (B_d)}$ 
 does not currently impose further constraints on the constants $G^{f_1f_2}_{X}$.

\section{Summary}
\label{sec:summary}
By comparing the inclusive decay rate
$\Gamma(B^+\to \psi  \cal B  \cal M )$ to the
exclusive decay rate $\Gamma(B^{+} \to p^{+} \psi)$
and taking their ratio, we obtained a lower bound
for the branching ratio Br$( B^{+} \to p^{+} \psi)$
as a function of only the dark matter particle's
mass $m_\psi$, independent of any arbitrary
coupling constants. 
This lower bound excludes the possibility of the
mass $m_\psi$ being larger than 3 GeV given the
most up to date upper bounds from BABAR, and
motivates further experimental analysis of this
decay as the remaining valid regions in parameter
space are relatively small, and may be explored
soon with future data, either confirming or excluding the  $B$-Mesogenesis model.
\\
Examining the lifetime ratios, we found the effect
of the $B$-Mesogenesis scenario to be minimal,
excluding the lifetime ratio as a discriminating 
observable for this model unless significant improvements are made in the SM uncertainty. 
\\
 Another future improvement of our analysis could be the determination of the exclusive decays with LCSR using the $B$ meson distribution amplitude, instead of the nucleon distribution amplitude, as done so far. 
 
\section*{Acknowledgements}
We thank Alexander Khodjamirian for many insightful discussions, a carefully reading of the manuscript
and suggesting to study the ratio
given in Eq.~\eqref{eq:ratio_decayrates} and 
Miguel Escudero  for helpful explanations related to the $B$ mesogenesis model.  A. M. and Z.
W. would like to thank Maria Laura Piscopo
and Aleksey Rusov for help with the first steps in the foundations of the heavy quark expansion and inclusive decays.
Moreover, we would like to thank
 Anastasia Boushmelev for explaining the calculations and error analysis of the exclusive decays $B^+ \to p^+ \psi$ in
 Ref. \cite{Boushmelev:2023huu}
.
 \\
This project was supported by the Deutsche Forschungsgemeinschaft (DFG, German Research Foundation) under grant 396021762-TRR 257 and
the BMBF project ``Theoretische Methoden für LHCb und Belle II''  (Förderkennzeichen 05H21PSCLA/ErUM-FSP T04).

\appendix

\section{\boldmath Numerical Inputs of the Constants $G^{f_1f_2}_{X}$}
\label{app:A}
In Tables \ref{Constants1/3} and \ref{Constants2/3}, we define and list the maximum values for the constants $G^{f_1f_2}_{X}$ introduced in Eq.~\eqref{eq:T-NP}, as estimated in~\cite{Alonso-Alvarez:2021qfd}.
\begin{longtable}{|>{\centering\arraybackslash}p{4.5cm} | >{\centering\arraybackslash}p{4cm} | >{\centering\arraybackslash}p{4cm}|}
    \caption{$\mathcal{H}_{-1/3}$ Constants and their Max values, taken from Ref.~\cite{Alonso-Alvarez:2021qfd}.}
    \label{Constants1/3}\\
    \hline
    Parameter & Max Value $\text{(GeV)}^{-4}$& Source \\
    \hline\hline
    \endfirsthead
    \multicolumn{3}{c}%
    {{\bfseries \tablename\ \thetable{} -- continued from previous page}} \\
    \hline
    Parameter & Max Value $\text{(GeV)}^{-4}$& Source \\
    \hline\hline
    \endhead
    \hline
    \multicolumn{3}{|r|}{{Continued on the next page}} \\
    \hline
    \endfoot
    \hline
    \endlastfoot
    $G^{\psi u}_{III}= \dfrac{y_{ub}y_{\psi d}}{M_{Y}^{2}}\dfrac{y^{*}_{ub}y^{*}_{\psi d}}{M_{Y}^{2}}$& $1\times 10^{-13}$ & Table II \\
    \hline
    $G^{\psi c}_{III}= \dfrac{y_{cb}y_{\psi d}}{M_{Y}^{2}}\dfrac{y^{*}_{cb}y^{*}_{\psi d}}{M_{Y}^{2}}$ & $1.5\times 10^{-12}$  &  Table II \\
    \hline
    $G^{u \psi}_{III}= \dfrac{y_{ud}y_{\psi b}}{M_{Y}^{2}}\dfrac{y^{*}_{ud}y^{*}_{\psi b}}{M_{Y}^{2}}$ & $4.7 \times 10^{-15}$ & Table II\\
    \hline
    $G^{c \psi}_{III}= \dfrac{y_{cd}y_{\psi b}}{M_{Y}^{2}}\dfrac{y^{*}_{cd}y^{*}_{\psi b}}{M_{Y}^{2}}$ &  $1.6 \times 10^{-13}$ &Table II\\
    \hline
    $G^{\psi u}_{II}= \dfrac{y_{ub}y_{\psi d}}{M_{Y}^{2}}\dfrac{y^{*}_{ud}y^{*}_{\psi b}}{M_{Y}^{2}}$ & $2.2 \times 10^{-14}$ & Table II \\
    \hline
    $G^{\psi c}_{II}= \dfrac{y_{cb}y_{\psi d}}{M_{Y}^{2}}\dfrac{y^{*}_{cd}y^{*}_{\psi b}}{M_{Y}^{2}}$& $4.9 \times 10^{-13}$ & Table II \\
    \hline
    $G^{u \psi}_{II}= \dfrac{y_{ud}y_{\psi b}}{M_{Y}^{2}}\dfrac{y^{*}_{ub}y^{*}_{\psi d}}{M_{Y}^{2}}$ & $2.2 \times 10^{-14}$ & Table II\\
    \hline
    $G^{c \psi}_{II}= \dfrac{y_{cd}y_{\psi b}}{M_{Y}^{2}}\dfrac{y^{*}_{cb}y^{*}_{\psi d}}{M_{Y}^{2}}$ &  $4.9 \times 10^{-13}$ & Table II \\
    \hline
    $G^{\psi \psi}_{I}= \dfrac{y_{\psi d}y^{*}_{\psi b}}{M_{Y}^{2}}\dfrac{y_{\psi b}y^{*}_{\psi d}}{M_{Y}^{2}}$  & $1.7 \times 10^{-15}$ & Eq. (43j), Eq. (41a) \\
    \hline
    $G^{u u}_{I}= \dfrac{y_{u d}y^{*}_{u b}}{M_{Y}^{2}}\dfrac{y_{u b}y^{*}_{u d}}{M_{Y}^{2}}$ &$1.7 \times 10^{-15}$ & Eq. (43f), Eq. (41a) \\
    \hline
    $G^{c c}_{I}= \dfrac{y_{c d}y^{*}_{c b}}{M_{Y}^{2}}\dfrac{y_{c b}y^{*}_{c d}}{M_{Y}^{2}}$  &$1.7 \times 10^{-15}$& Eq. (43f), Eq. (41a) \\
    \hline
    $G^{c u}_{I}= \dfrac{y_{u d}y^{*}_{c b}}{M_{Y}^{2}}\dfrac{y_{c b}y^{*}_{u d}}{M_{Y}^{2}}$  & $8.5 \times 10^{-16}$ & Eq. (43e), Eq. (41a) \\
    \hline
    $G^{uc}_{I}= \dfrac{y_{c d}y^{*}_{u b}}{M_{Y}^{2}}\dfrac{y_{u b}y^{*}_{c d}}{M_{Y}^{2}}$& $8.5 \times 10^{-16}$ & Eq. (43e), Eq. (41a) \\
    \hline
    $G^{s \psi}_{IV}= \dfrac{y_{\psi s}y_{u b}}{M_{Y}^{2}}\dfrac{y^{*}_{u b}y^{*}_{\psi s}}{M_{Y}^{2}}$ & $1 \times 10^{-13}$ & Table II \\
     \hline
    $G^{d \psi}_{IV}= \dfrac{y_{\psi d}y_{u b}}{M_{Y}^{2}}\dfrac{y^{*}_{u b}y^{*}_{\psi d}}{M_{Y}^{2}}$ & $1 \times 10^{-13}$ & Table II \\
    \hline
    $G^{\psi s}_{III}= \dfrac{y_{\psi b}y_{u s}}{M_{Y}^{2}}\dfrac{y^{*}_{u s}y^{*}_{\psi b}}{M_{Y}^{2}}$  & $3.7 \times 10^{-14}$ & Table II \\
    \hline
    $G^{\psi d}_{III}= \dfrac{y_{\psi b}y_{u d}}{M_{Y}^{2}}\dfrac{y^{*}_{u d}y^{*}_{\psi b}}{M_{Y}^{2}}$  & $4.7 \times 10^{-15}$ & Table II \\
    \hline
    $G^{\psi s}_{V}= \dfrac{y_{\psi s}y_{u b}}{M_{Y}^{2}}\dfrac{y^{*}_{u s}y^{*}_{\psi b}}{M_{Y}^{2}}$  & $6.3 \times 10^{-14}$ & Table II \\
    \hline
    $G^{\psi d}_{V}= \dfrac{y_{\psi d}y_{u b}}{M_{Y}^{2}}\dfrac{y^{*}_{u d}y^{*}_{\psi b}}{M_{Y}^{2}}$  & $2.2 \times 10^{-14}$ & Table II \\
    \hline
    $G^{\psi s}_{VI}= \dfrac{y_{\psi b}y_{u s}}{M_{Y}^{2}}\dfrac{y^{*}_{u b}y^{*}_{\psi s}}{M_{Y}^{2}}$  & $6.3 \times 10^{-14}$ & Table II \\
    \hline
    $G^{\psi d}_{VI}= \dfrac{y_{\psi b}y_{u d}}{M_{Y}^{2}}\dfrac{y^{*}_{u b}y^{*}_{\psi d}}{M_{Y}^{2}}$  & $2.2 \times 10^{-14}$ & Table II \\
    \hline
    $G^{cs}_{IV}= \dfrac{y_{c s}y^{*}_{u b}}{M_{Y}^{2}}\dfrac{y_{u b}y^{*}_{c s}}{M_{Y}^{2}}$ & $8.5 \times 10^{-15}$ & Eq. (43g), Eq. (41a) \\
    \hline
    $G^{us}_{IV}= \dfrac{y_{u s}y^{*}_{u b}}{M_{Y}^{2}}\dfrac{y_{u b}y^{*}_{u s}}{M_{Y}^{2}}$  &$8.5 \times 10^{-15}$& Eq. (43g), Eq. (41a) \\
     \hline
    $G^{cd}_{IV}= \dfrac{y_{c d}y^{*}_{u b}}{M_{Y}^{2}}\dfrac{y_{u b}y^{*}_{c d}}{M_{Y}^{2}}$ & $8.5 \times 10^{-16}$ & Eq. (43e), Eq. (41a) \\
    \hline
    $G^{ud}_{IV}= \dfrac{y_{u d}y^{*}_{u b}}{M_{Y}^{2}}\dfrac{y_{u b}y^{*}_{u d}}{M_{Y}^{2}}$ & $1.7 \times 10^{-15}$ & Eq. (43f), Eq. (41a) \\
     \hline
    $G^{cs}_{I}= \dfrac{y_{u s}y^{*}_{c b}}{M_{Y}^{2}}\dfrac{y_{c b}y^{*}_{u s}}{M_{Y}^{2}}$  & $8.5\times 10^{-15}$ & Eq. (43g), Eq. (41a)\\
     \hline
    $G^{us}_{I}= \dfrac{y_{u s}y^{*}_{u b}}{M_{Y}^{2}}\dfrac{y_{u b}y^{*}_{u s}}{M_{Y}^{2}}$  & $8.5\times 10^{-15}$ & Eq. (43g), Eq. (41a)\\
    \hline
    $G^{cd}_{I}= \dfrac{y_{u d}y^{*}_{c b}}{M_{Y}^{2}}\dfrac{y_{c b}y^{*}_{u d}}{M_{Y}^{2}}$  & $8.5\times 10^{-16}$ & Eq. (43e), Eq. (41a)\\
    \hline
    $G^{ud}_{I}= \dfrac{y_{u d}y^{*}_{u b}}{M_{Y}^{2}}\dfrac{y_{u b}y^{*}_{u d}}{M_{Y}^{2}}$  & $1.7\times 10^{-15}$ & Eq. (43f), Eq. (41a)\\
    
\end{longtable}
\begin{table}[h]
    \centering
    \renewcommand{\arraystretch}{2}
    \caption{ $\mathcal{H}_{2/3}$ Constants and their Max values, taken from Ref.~\cite{Alonso-Alvarez:2021qfd}.}
    \label{Constants2/3}
    \begin{tabular} { |>{\centering\arraybackslash}p{4.5cm} | >{\centering\arraybackslash}p{4cm} | >{\centering\arraybackslash}p{4cm}|}
        \hline
        Parameter & Max Value $\text{(GeV)}^{-4}$& Source \\
        \hline\hline
        $G^{u \psi}_{IV}= \dfrac{y_{bd}y_{\psi u}}{M_{Y}^{2}}\dfrac{y^{*}_{bd}y^{*}_{\psi u}}{M_{Y}^{2}}$& $2.5\times 10^{-13}$ & Table II \\
        \hline
        $G^{c \psi}_{IV}= \dfrac{y_{bd}y_{\psi c}}{M_{Y}^{2}}\dfrac{y^{*}_{bd}y^{*}_{\psi c}}{M_{Y}^{2}}$ & $2.5\times 10^{-13}$  &  Table II \\
        \hline
        $G^{d s}_{IV}= \dfrac{y_{ds}y^{*}_{b d}}{M_{Y}^{2}}\dfrac{y_{bd}y^{*}_{d s}}{M_{Y}^{2}}$ &  $8.5 \times 10^{-15}$  & Eq. (44d), Eq. (41a)\\
       \hline
        $G^{d s}_{I}= \dfrac{y_{ds}y^{*}_{b d}}{M_{Y}^{2}}\dfrac{y_{bd}y^{*}_{d s}}{M_{Y}^{2}}$ &  $8.5 \times 10^{-15}$  & Eq. (44d), Eq. (41a)\\
        \hline
        $G^{s s}_{I}= \dfrac{y_{ds}y^{*}_{b s}}{M_{Y}^{2}}\dfrac{y_{bs}y^{*}_{d s}}{M_{Y}^{2}}$ &  $1.7 \times 10^{-15}$  & Eq. (44c), Eq. (41a)\\
         \hline
        $G^{d \psi}_{III}= \dfrac{y_{bd}y_{\psi u}}{M_{Y}^{2}}\dfrac{y^{*}_{bd}y^{*}_{\psi u}}{M_{Y}^{2}}$ & $2.5\times 10^{-13}$ & Table II \\
        \hline
        $G^{s \psi}_{III}= \dfrac{y_{bs}y_{\psi u}}{M_{Y}^{2}}\dfrac{y^{*}_{bs}y^{*}_{\psi u}}{M_{Y}^{2}}$ & $1.2\times 10^{-12}$ & Table II \\
        \hline
    \end{tabular}
\end{table}
\clearpage
\noindent
\section{Supplementary plots}
\label{supplement}
\begin{figure}[H]
    \includegraphics[scale=0.8]{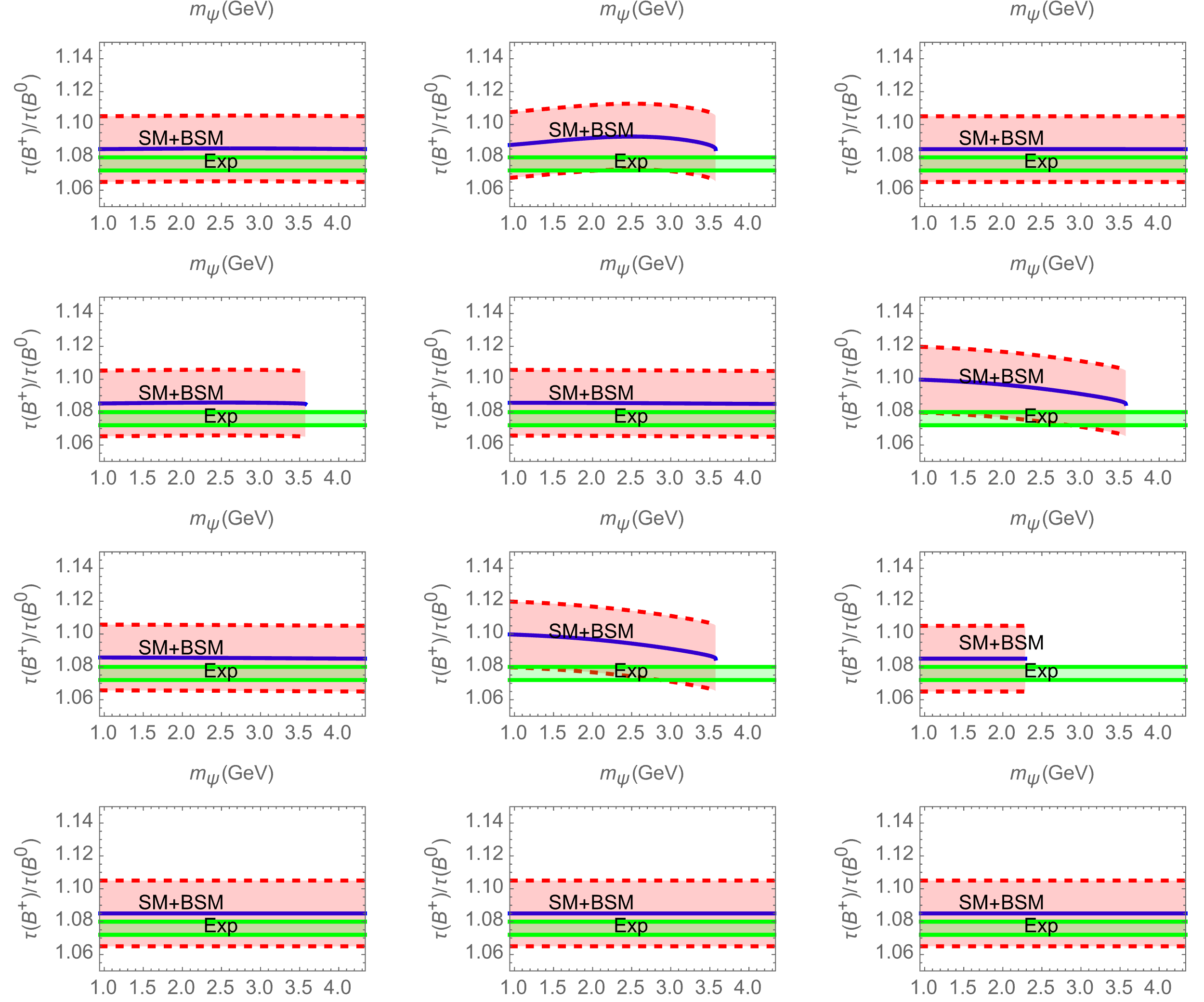}
    \caption{The lifetime ratio ${\tau (B^+)}/{\tau (B_d)}$ for $\mathcal{H}_{-1/3}$ as a function of $m_{\psi}$, using the upper limits of one the constants $G^{q1q2}_{X}$ and setting the others to zero. The red region shows the SM error, whereas the experimental value is shown in green. The non-vanishing constants  from up to down and left to right are $G^{\psi u}_{III}, G^{\psi c}_{III},\, G^{u \psi}_{III}, \, G^{c \psi }_{III}, G^{\psi u}_{II}, \, G^{\psi c}_{II}, \, G^{u \psi }_{II},\, G^{c \psi}_{II},\, G^{\psi \psi}_{I},\, G^{u u}_{I} = G^{c c}_{I},\, G^{c u}_{I},\, G^{u c}_{I} $ }
    \label{fig:Lifetime1/3-BdAllPlots}
\end{figure}

\begin{figure}[H]
    \includegraphics[scale=0.8]{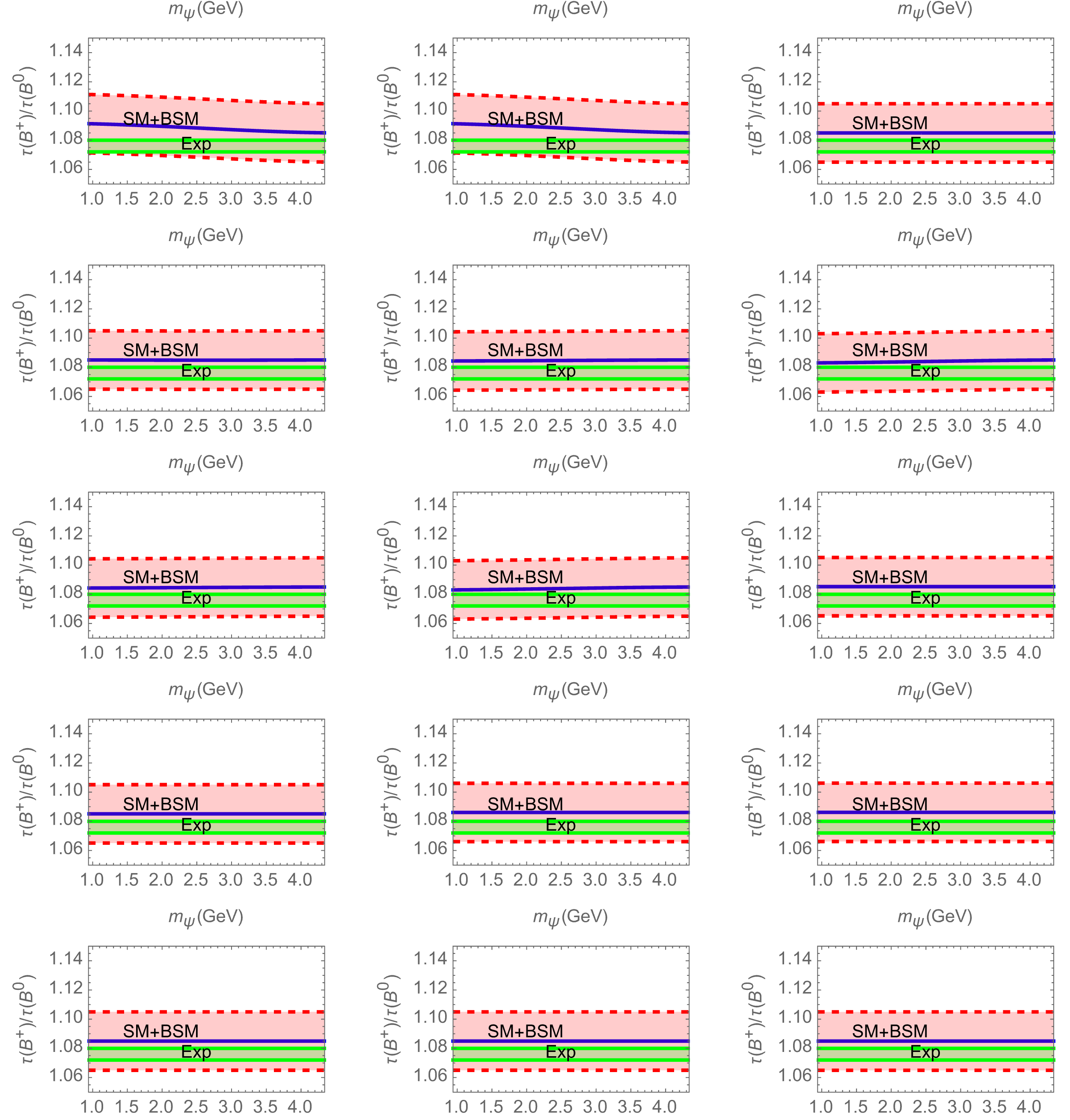}
    \caption{The lifetime ratio ${\tau (B^+)}/{\tau (B_d)}$ for $\mathcal{H}_{-1/3}$ as a function of $m_{\psi}$, using the upper limits of one the constants $G^{f_1f_2}_{X}$ and setting the others to zero. The red region shows the SM error, whereas the experimental value is shown in green. The non-vanishing constants  from up to down and left to right are $G^{\psi d}_{IV}, G^{\psi s}_{IV},\, G^{\psi d}_{III}, \, G^{\psi s}_{III}, G^{\psi d}_{V}, \, G^{\psi s}_{V}, \, G^{\psi d}_{VI},\, G^{\psi s}_{VI}, \, G^{u d}_{IV}, \, G^{c d}_{IV}, \, G^{s c}_{I}, \, G^{s u}_{IV}, \, G^{d u}_{I}, \, G^{d c}_{I}, \, G^{s c}_{I}, \, G^{s u}_{I}.$ }
    \label{fig:Lifetime1/3-BPlusAllPlots}
\end{figure}

In Figures~\ref{fig:Lifetime1/3-BdAllPlots}  and~\ref{fig:Lifetime1/3-BPlusAllPlots}, we show the lifetime ratio ${\tau (B^+)}/{\tau (B_d)}$ as a function of the dark sector mass $m_{\psi}$ when only one constant $G^{f_1f_2}_{X}$ is turned on to their maximum value and the others are set to zero. The maximum values for the different constants $G^{f_1f_2}_{X}$ are provided in Tables \ref{Constants1/3} and \ref{Constants2/3}.  
\begin{figure}[H]
    \includegraphics[scale=0.8]{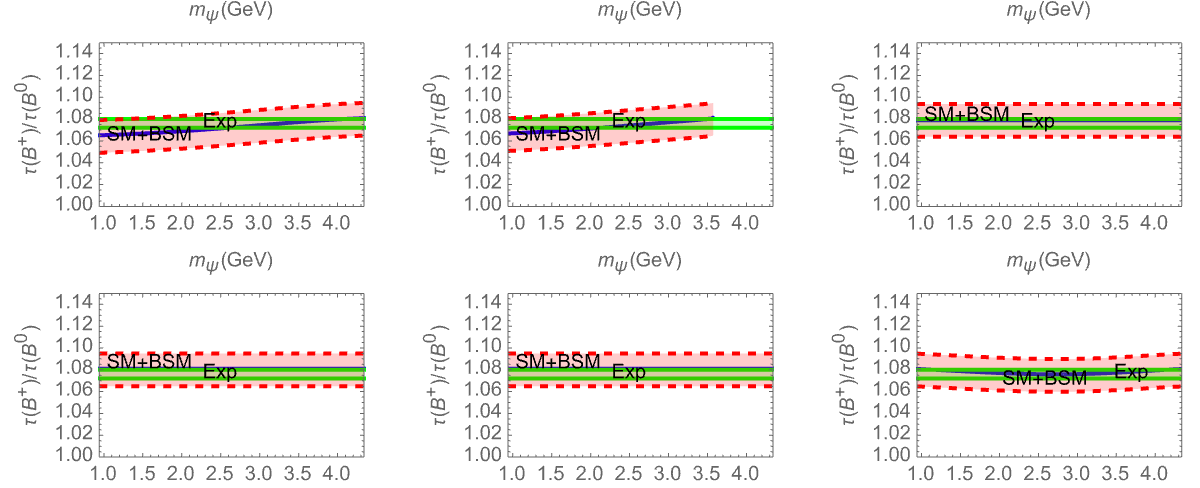}
    \caption{The lifetime ratio ${\tau (B^+)}/{\tau (B_d)}$ for $\mathcal{H}_{2/3}$ as a function of $m_{\psi}$, using the upper limits of one the constants $G^{f_1f_2}_{X}$ and setting the others to zero. The red region shows the SM error, whereas the experimental value is shown in green. The non-vanishing constants  from up to down and left to right are $G^{u \psi}_{IV}, G^{c \psi}_{IV},\, G^{d s}_{IV}, \, G^{d s}_{I}, G^{s s}_{I}= \, G^{d \psi}_{III}, \, G^{s \psi}_{III},$}
    \label{fig:Lifetime2/3AllPlots}
\end{figure}

\clearpage

\bibliographystyle{JHEP}
\bibliography{References}

\end{document}